\documentclass[aip,rsi,reprint]{revtex4-1}

\usepackage{graphicx}%
\usepackage{color}%
\usepackage{dcolumn}%
\usepackage{bm}%
\usepackage[]{units}

\begin{document}

\title{Magnetic-film atom chip with \unit[10]{$\mu$m} period lattices of microtraps for quantum information science with Rydberg atoms}

\author{V.Y.F.~Leung}
\affiliation {Van der Waals-Zeeman Institute, University of Amsterdam, \\
Science Park 904, PO Box 94485, 1090 GL Amsterdam, The Netherlands}
\affiliation {Complex Photonic Systems (COPS), MESA+ Institute for Nanotechnology, \\
University of Twente, PO Box 217, 7500 AE Enschede, The Netherlands}

\author{D.R.M.~Pijn}
\affiliation{Van der Waals-Zeeman Institute, University of Amsterdam, \\
Science Park 904, PO Box 94485, 1090 GL Amsterdam, The Netherlands}
\author{H.~Schlatter}
\affiliation{Van der Waals-Zeeman Institute, University of Amsterdam, \\
Science Park 904, PO Box 94485, 1090 GL Amsterdam, The Netherlands}
\author{L.~Torralbo-Campo}
\affiliation{Van der Waals-Zeeman Institute, University of Amsterdam, \\
Science Park 904, PO Box 94485, 1090 GL Amsterdam, The Netherlands}
\author{A.L.~La Rooij}
\affiliation{Van der Waals-Zeeman Institute, University of Amsterdam, \\
Science Park 904, PO Box 94485, 1090 GL Amsterdam, The Netherlands}
\author{G.B.~Mulder}
\affiliation{Van der Waals-Zeeman Institute, University of Amsterdam, \\
Science Park 904, PO Box 94485, 1090 GL Amsterdam, The Netherlands}
\author{J.~Naber}
\affiliation{Van der Waals-Zeeman Institute, University of Amsterdam, \\
Science Park 904, PO Box 94485, 1090 GL Amsterdam, The Netherlands}
\author{M.L.~Soudijn}
\affiliation{Van der Waals-Zeeman Institute, University of Amsterdam, \\
Science Park 904, PO Box 94485, 1090 GL Amsterdam, The Netherlands}
\author{A.~Tauschinsky}
\affiliation{Van der Waals-Zeeman Institute, University of Amsterdam, \\
Science Park 904, PO Box 94485, 1090 GL Amsterdam, The Netherlands}
\author{C.~Abarbanel}
\affiliation {Ilse Katz Institute for Nanoscale Science and Technology,
Ben-Gurion University of the Negev, Be'er Sheva 84105, Israel}
\author{B.~Hadad}
\affiliation {Ilse Katz Institute for Nanoscale Science and Technology,
Ben-Gurion University of the Negev, Be'er Sheva 84105, Israel}
\author{E.~Golan}
\affiliation {Ilse Katz Institute for Nanoscale Science and Technology,
Ben-Gurion University of the Negev, Be'er Sheva 84105, Israel}
\author{R.~Folman}
\affiliation{Department of Physics and Ilse Katz Institute for Nanoscale Science and Technology,
Ben-Gurion University of the Negev, Be'er Sheva 84105, Israel}

\author{R.J.C.~Spreeuw}

\email{r.j.c.spreeuw@uva.nl}
\affiliation{Van der Waals-Zeeman Institute, University of Amsterdam, \\
Science Park 904, PO Box 94485, 1090 GL Amsterdam, The Netherlands}
\homepage {http://www.science.uva.nl/research/aplp/}
\date{\today}

\begin{abstract}

We describe the fabrication and construction of a setup for creating lattices of magnetic microtraps for ultracold atoms on an atom chip. The lattice is defined by lithographic patterning of a permanent magnetic film. Patterned magnetic-film atom chips enable a large variety of trapping geometries over a wide range of length scales. We demonstrate an atom chip with a lattice constant of \unit[10]{$\mu$m}, suitable for experiments in quantum information science employing the interaction between atoms in highly-excited Rydberg energy levels.  The active trapping region contains lattice regions with square and hexagonal symmetry, with the two regions joined at an interface.
A structure of macroscopic wires, cut out of a silver foil, was mounted under the  atom chip in order to load ultracold $^{87}$Rb atoms into the microtraps. We demonstrate loading of atoms into the square and hexagonal lattice sections simultaneously and show resolved imaging of individual lattice sites. Magnetic-film lattices on atom chips provide a versatile platform for experiments with ultracold atoms, in particular for quantum information science and quantum simulation.

\end{abstract}
\maketitle

\section{Introduction}

The last decade has witnessed the development of lattice structures for trapped ultracold neutral atoms or ions, where they can be coherently controlled and manipulated. These lattices have already been very successful as platforms to simulate a large variety of quantum phenomena, including quantum many-body physics~\cite{GreManBlo02,Endres2011Observation,Greif2013ShortRange,Bloch2012Quantum,Simon2011Quantum,Seidelin2006,Blatt2008Entangled,Islam2011Onset,Lanyon2011Universal, WelBauAba11}.

For neutral atoms, optical lattices have been the norm, with lattice parameters of about \unit[0.5]{$\mu$m} and interaction among neighbouring sites provided by tunneling. 
Lattices with larger lattice parameters, in the range of \unit[5-10]{$\mu$m}, are now attracting increasing interest \cite{Piotrowicz2013Two-dimensional,Nogrette2014Singleatom}. This longer length scale is well suited to implement tunable,  switchable interactions among individually addressable sites, by exciting atoms into highly excited  Rydberg levels. This approach is presently subject of intense theoretical and experimental investigation \cite{JakCirLuk00,LukFleZol01,WeiMulBuc10,Saffman2010Quantum,WilGaeBro10,IseUrbSaf10}. 

We recently demonstrated an alternative technique to create two-dimensional neutral-atom lattices, based on a patterned film of magnetic iron-platinum (FePt) alloy on an atom chip \cite{GerWhiFer07,WhiGerSpr09}. This  hybrid system combines the advantages of lithography techniques with cold atom experiments.
Magnetic lattices offer unique opportunities that are in  many ways complementary to those offered by the optical variety\cite{SinVolAku08,LlorenteGarcia2010Experiments,Boyd2007Atom}. They can be used to extend the range of length scales to both larger and smaller lattice parameters, each giving access to very different regimes of quantum simulation \cite{LeuTauSpr11}. While smaller ($\sim\unit[100]{nm}$) length scales could push Hubbard models into  parameter regimes beyond what is now achievable using optical lattices \cite{Gullans2012Nanoplasmonic,Romero-Isart2013}, in this paper we concentrate on the longer length scale.

Here we present a detailed description of an improved setup for magnetic-film atom chip lattice experiments, utilizing  \unit[10]{$\mu$m} lattice spacings, geared toward the Rydberg route described above. This atom chip is an important advance towards the application of chip-based magnetic lattices as a quantum information platform where interactions will be mediated by exciting atoms to Rydberg levels. The spacing of \unit[10]{$\mu$m} is within reach of dipole-dipole interactions between atoms in highly-excited Rydberg levels~\cite{DitKoeHer08,Saffman2010Quantum,WilGaeBro10,IseUrbSaf10}. At the same time the lattice spacing is large enough to easily resolve individual traps, providing optical  access at will, using off-the-shelf optics. 
Since the atom-to-chip trapping distance scales with the lattice period, these chip based traps also provide access to measurements of the interaction of (Rydberg) atoms with the surface at the micron scale \cite{TauThiSpr10,Abel2011Electrometry,Kubler2010Coherent,Hattermann2012Detrimental}.

Magnetic lattices offer virtually unlimited flexibility in the design of lattice patterns, including non-periodic patterns and controlled amounts of disorder. 
Regions of different  functionalities, or lattices with different length scales or geometries can be combined in a single pattern. 
As an example of the design flexibility that magnetic lattices have to offer, we demonstrate the loading of atoms into a square and a hexagonal lattice that are joined at an interface.

The paper is structured as follows. In Sec.\ II we present the microfabrication procedure used for the fabrication of this atom chip. In Sec.\ III we describe the technical aspects of the chip mount assembly. In Sec.\ IV we present our first experimental results and demonstrate the loading of atoms into the microtraps.

\section{Micro-fabrication of the Magnetic Film Atom Chip}\label{section2}

The atom chip is based on iron platinum, a hard magnetic material, so that the magnetization is robust against externally applied magnetic fields that are small compared to the coercivity \cite{XinBarGer07}.  The FePt films were supplied by Hitachi Research San Jose and have already been used in the fabrication of our previous generation of atom chips \cite{GerWhiFer07}.
The films have high remanent magnetization ($M_r=\unit[670]{kA/m}$), coercivity ($\mu_0 H_c=\unit[0.95]{T}$), and Curie temperature (estimated $T_C\gtrsim\unit[450]{^\circ C}$)\cite{Barmak2005}. Using scanning electron microscopy (SEM) we find that the size of individual grains of FePt within our samples is typically \unit[40]{nm}.  This  grain size currently poses no limitation for the \unit[10]{$\mu$m} period lattices presented here. 
The atom chip consists of a \unit[200]{nm} thick FePt film deposited on a \unit[330]{$\mu$m} thick silicon substrate,  $15\times$\unit[20]{mm$^2$} in size.

The magnetic lattice pattern has been designed using an optimization algorithm for creating traps in arbitrary lattice configurations, while optimizing for maximum trap stiffness \cite{SchSprWhi10}. The optimization typically yields a binary pattern containing pixels with either zero or maximal magnetization. Such patterns can easily be produced by etching away the pixels with zero magnetization. The versatility of this technique allows us to optimize both the square and hexagonal lattice geometries used in our experiment. These two designs are then connected together at an interface. With an eye on the intended application as a quantum information platform, an obvious difference of the two geometries is the number of nearest neighbors of each lattice site.

\subsection{Patterning, Etching, and Planarization}

\begin{figure}
    \centering
    \includegraphics[width=3.5in]{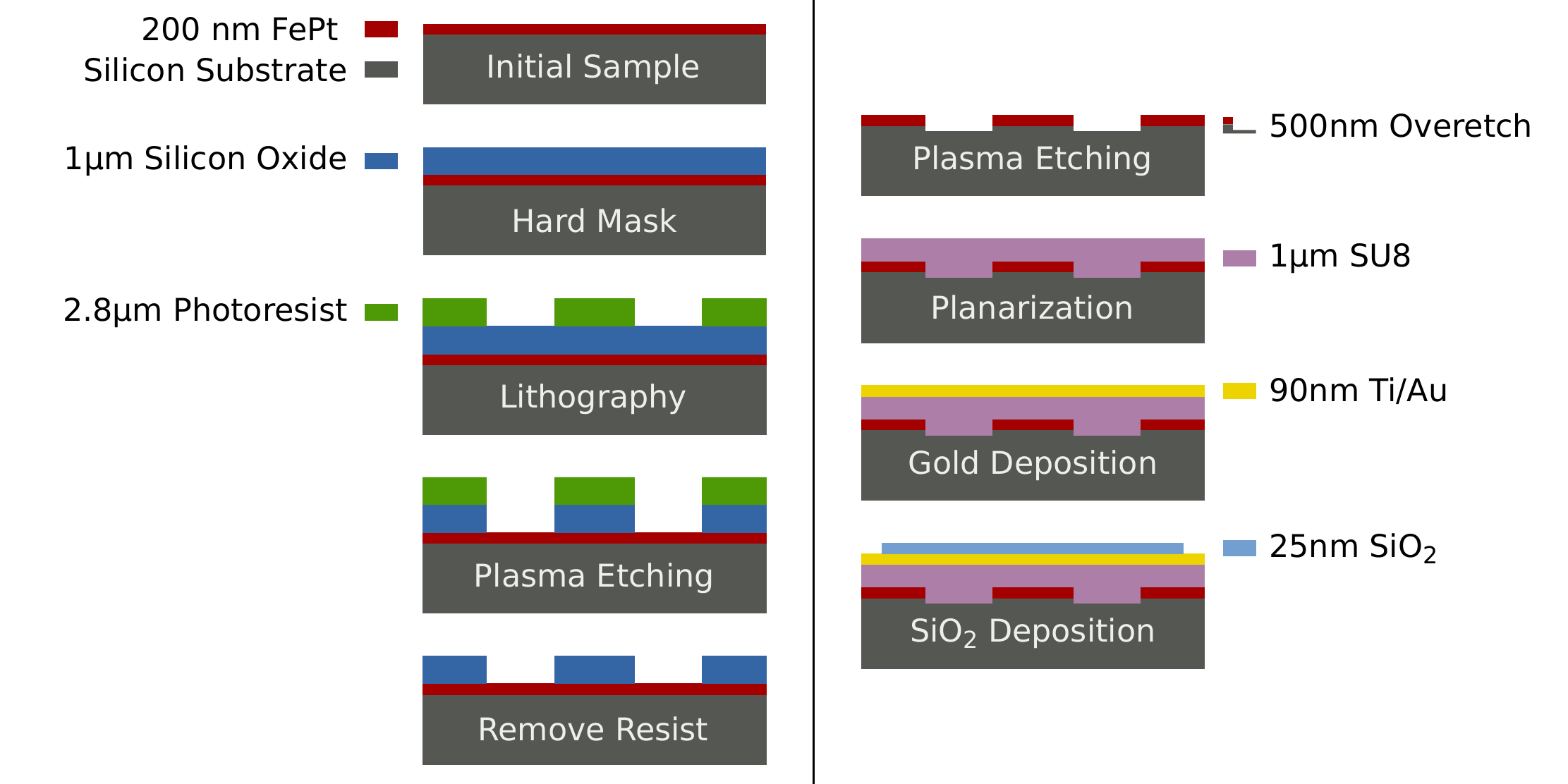}
    \caption{Processing steps in the chip fabrication sequence, see text.}
    \label{process}
\end{figure}

The design pattern is transferred to the FePt film by UV-optical lithography at the fabrication facility of Ben-Gurion University of the Negev, Israel.
The lithography procedure is shown in Fig.\ref{process}.  First, a lithographic mask is made by depositing a layer of \unit[1]{$\mu$m} of silicon oxide onto the FePt, followed by \unit[2.8]{$\mu$m} of photo-resist. After the exposure and development of the resist, the silicon oxide is plasma-etched.  The remaining photo-resist is removed by wet etching. Using the silicon oxide as the mask, the FePt is then etched by direct argon ion etching to complete the pattern transfer.

The chip surface also acts as a mirror for the magneto-optical trap (MOT)\cite{RaaPrenCab87,ReiHanHan99} and for the imaging beams.  Therefore optical flatness is required to reduce the formation of interference fringes resulting from light reflecting off the etched and non-etched regions of the chip.  As the height difference between etched and non-etched regions (including  overetch) is similar to the wavelength of the light (\unit[780]{nm}) used for cooling and imaging of the atoms, interference effects are particularly relevant. The etched surface is therefore planarized with the polymer SU8, commonly used as a photo-resist.  After deposition, the chip is baked at \unit[105]{$^\circ$C} for 24 hours, to increase the hardness of the SU8 and to limit its outgassing in ultra-high vacuum. The total thickness of the deposited SU8 is \unit[1]{$\mu$m} with a flatness of less than \unit[9]{nm}, as measured by atomic force microscopy (AFM).

\begin{figure}
    \centering
    \includegraphics[width=3.5in]{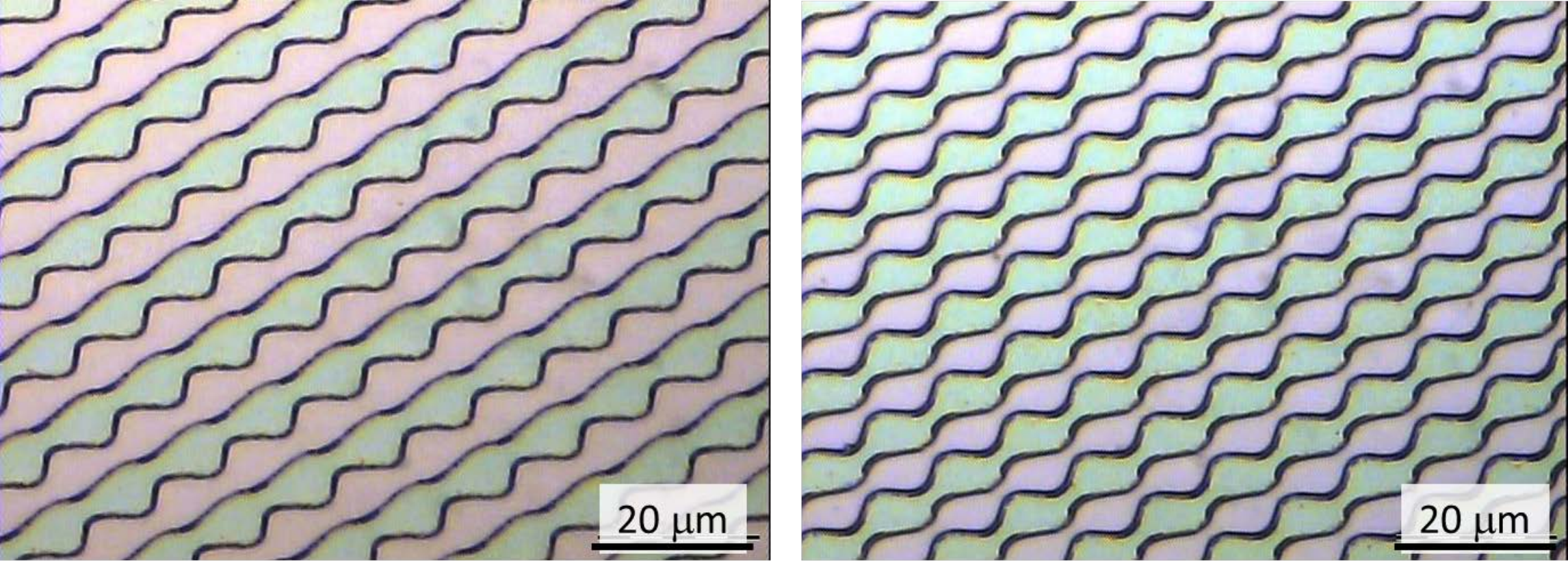}
    \caption{Patterns on the chip after the etching process. The alternating strips are etched and non-etched regions. The pattern on the left generates a magnetic lattice with square translational symmetry. The lattice pattern on the right has hexagonal (equilateral triangular) symmetry. }
    \label{etchedpatterns}
\end{figure}

\subsection{Reflective and protective coatings}\label{layerCoating}

After planarization a \unit[90]{nm} thick gold coating is deposited, giving the surface the necessary reflective properties for the mirror MOT. In previous experiments we found that the non-uniform adsorption of rubidium atoms onto the gold chip surface led to stray electric fields. These could be observed as energy shifts of Rydberg levels by electromagnetically induced transparency (EIT) spectroscopy \cite{TauThiSpr10}.
  To prevent the direct exposure of the gold surface to rubidium vapor, an extra layer of \unit[25]{nm} of quartz (SiO$_{2}$) is deposited on top of the gold surface \cite{BoFenMin09}. Rubidium atoms adsorb less strongly to quartz than to gold and are more easily removed via light-induced atomic desorption (LIAD) with blue or ultraviolet light \cite{GozBer93, KleArl06}. The corners of the chip are left uncoated with quartz so that the gold surface can be electrically grounded to the mount.
As the final step in the preparation of the atom chip, the FePt layer is magnetized  by applying an external magnetic field of 5 T in the out-of-plane direction.

\section{Design and construction of the atom chip mount}\label{section3}

The chip is clamped on top of a construction of silver wires. These wires support electrical currents and serve to magnetically trap atoms before they are loaded into the chip lattice. The silver wire pattern is glued on top of a copper mounting structure for mechanical stability and to transport heat away from the chip. The glue also acts as an electrical insulator.  The entire construction is mounted in ultrahigh vacuum. This section describes the various components.

\subsection{Supporting wire pattern}

\begin{figure}[h]
    \centering
    \includegraphics[width=3.4in] {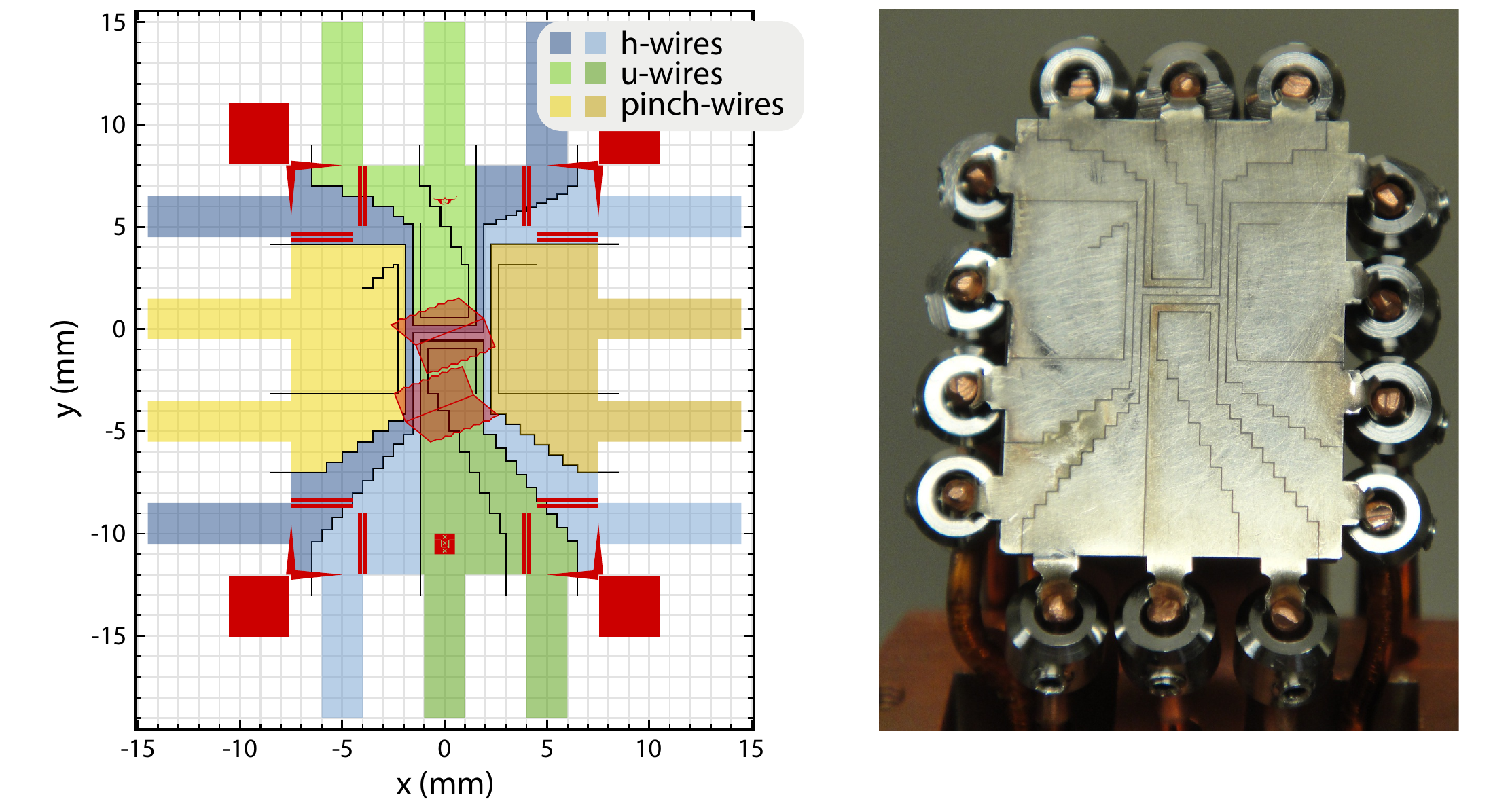}
    \caption{Silver foil: (Left) schematic drawing with Z- and U-wires for atom trapping. Overlayed in red-brown are the two regions of permanent magnet patterns.
    (Right) image of the silver foil, glued onto its copper base, and fully connected. }
    \label{silverfoil}
\end{figure}

The wire pattern design as shown in Fig. \ref{silverfoil} features two opposing ``h"-shaped silver wires.  Each ``h"-wire can be used in either a ``U" or a ``Z" configuration.  The U-sections in the h-wires are complemented by an additional set of U-wires for greater current capacity and more flexibility in positioning.
Sending current through the U-wire allows us to trap atoms in a MOT near the surface. The Z-wires, in combination with an external bias magnetic field, create an Ioffe-Pritchard (IP) type magnetic trap \cite{FolFruScht02,DuReiImi04,ReiHanHan99}.  This IP trap is used to load atoms and to position the initial cloud before loading the atoms into the final magnetic microtraps.  The Z-wires are centered beneath one magnetic lattice region out of the possible two regions, shown as a red-brown patch in Fig. \ref{silverfoil} [around $(x, y) = (0,0)$].  Having duplicate trapping regions gives us the possibility of switching the orientation of the chip or use the second trapping region, should the first region be damaged.

The outermost wires along the $y$-direction can be used as ``pinch" wires, providing extra axial confinement for the Z-wire trap. Unfortunately, the silver wires have not been useful for delivering radio-frequency (RF) magnetic fields to the atoms for forced evaporative cooling. This is due to the strong capacitive coupling between the wires and the copper base, effectively shorting the RF current before the power reaches the active part of the chip.

\begin{figure}[h]
    \centering
    \includegraphics[width=2.5in] {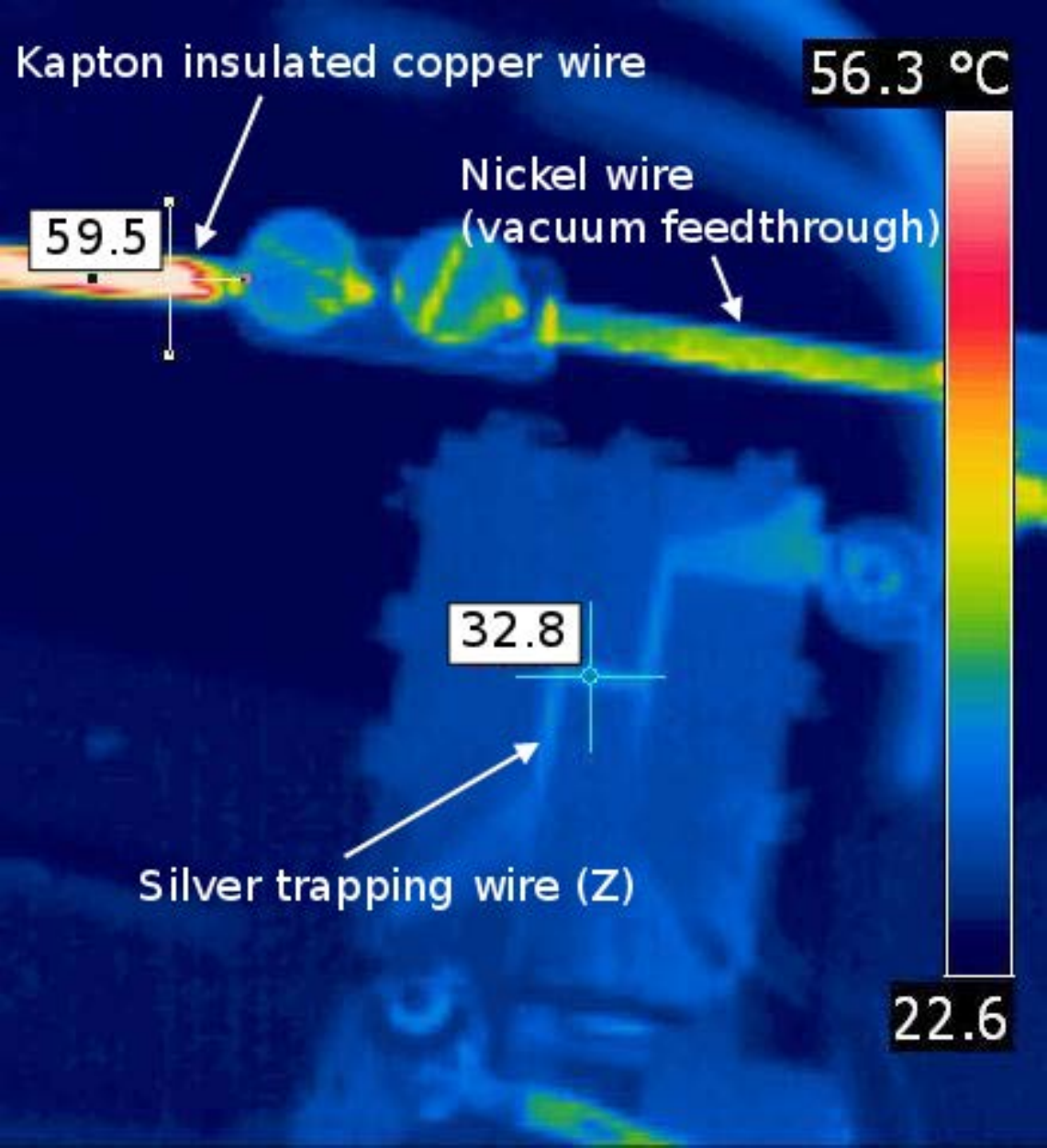}
    \caption{Heating bench test of the silver foil and a connector to the vacuum feedthrough. The lower part shows a thermal infrared image of one of the Z-wires while carrying 20 A. The upper part of the image shows one connector to the vacuum feedthrough. Note that inside the vacuum system these components are far away from each other. They have been arranged on the bench to appear in one picture.}
    \label{thermacam}
\end{figure}

The fabrication procedure of the silver wire pattern begins with the preparation of a \unit[250]{$\mu$m} thick sheet of tempered, flattened silver foil (Goodfellow).  This foil is first annealed in argon gas with 5\% H$_2$ at \unit[650]{$^\circ$C} and then reflattened at a pressure of \unit[10]{kN/cm$^2$}.

Spark erosion is used to cut the foil  into the trapping wire pattern shown in Fig.~\ref{silverfoil}.
For the spark erosion, AGIE EMT 1.1 LF equipment is used with a wire of \unit[50]{$\mu$m} diameter (sample courtesy of Microcut). During the cutting process the wires are held together in a frame, which remains in place until after the gluing stage.  After rinsing with ethanol, the trapping wires are argon plasma-etched in an AJA sputtering device and coated on the gluing side with \unit[$\sim$40]{nm} chromium to promote adhesion.

We use a low out-gassing, high thermally conducting, electrically insulating epoxy (Epotek H77S) to fill the gaps between the trapping wires and secure the silver foil to the copper base. The mixed H77S epoxy is degassed for 1 hour at \unit[40]{$^\circ$C} after which \unit[$\sim$0.05]{g} is applied to the chromium coated side of the trapping wires. After waiting until the gaps are filled, the trapping wire foil is carefully placed on the sandblasted and leveled copper base. The trapping foil is forced onto the copper base by capillary action, leaving only a monolayer of glue filler grains in-between the silver and the copper. The \unit[20]{$\mu$m} boron nitride (BN) filler grains of the H77S act both as an electrical insulator and a separator, keeping the trapping wires away from the copper base with minimal thermal resistance. The glue is hardened for 1 hour at \unit[150]{$^\circ$C}. Excess glue on the top of the foil is ground away.

\subsection{Mechanical construction}

The electrical connections to the silver wires are provided by 14 Kapton insulated copper wires leading to a 20-pin vacuum feedthrough. The current is limited to \unit[10]{A} per pin by the maximum loading of the vacuum feedthrough. IR imaging of the wires placed outside the vacuum reveals a \unit[$\sim$10]{K} temperature rise at  \unit[20]{A} (using two feedthrough pins in parallel), as shown in Fig. \ref{thermacam}. This test shows that the heat is mainly concentrated in the connections to the electrical feedthrough, where temperatures up to \unit[60]{$^\circ$C} have been observed. Using square wires cut out of silver foil allows us to run high currents as close as possible to the magnetic structure. This creates a very localized and high-gradient magnetic trap, with field gradients up to \unit[$\sim$15]{T/m} and usable radial trap frequencies up to \unit[$\sim$2]{kHz}.

The silver flaps extending from the trapping wires are connected to the vacuum feedthrough with \unit[1.5]{mm} diameter Kapton-insulated, oxygen-free-high-conductivity (OFHC) copper wires. To protect the silver joints, the silver flaps are pre-bent in order to prevent damage by thermal expansion, and the tips of the clamping screws are slightly rounded to minimize torsion. To allow a current of \unit[20]{A} to flow through the Z-shaped wires, two pins of the feedthrough are connected in parallel. The titanium socket for the rubidium dispenser is screwed onto the copper base keeping the dispenser just below the line of sight of the atom chip surface.

The chip is clamped directly onto the silver wires by four titanium springs, allowing for easy replacement of the chip. The chip surface is electrically grounded to the copper base via the titanium springs.

\begin{figure}[h]
    \centering
    \includegraphics[width=3.5in]{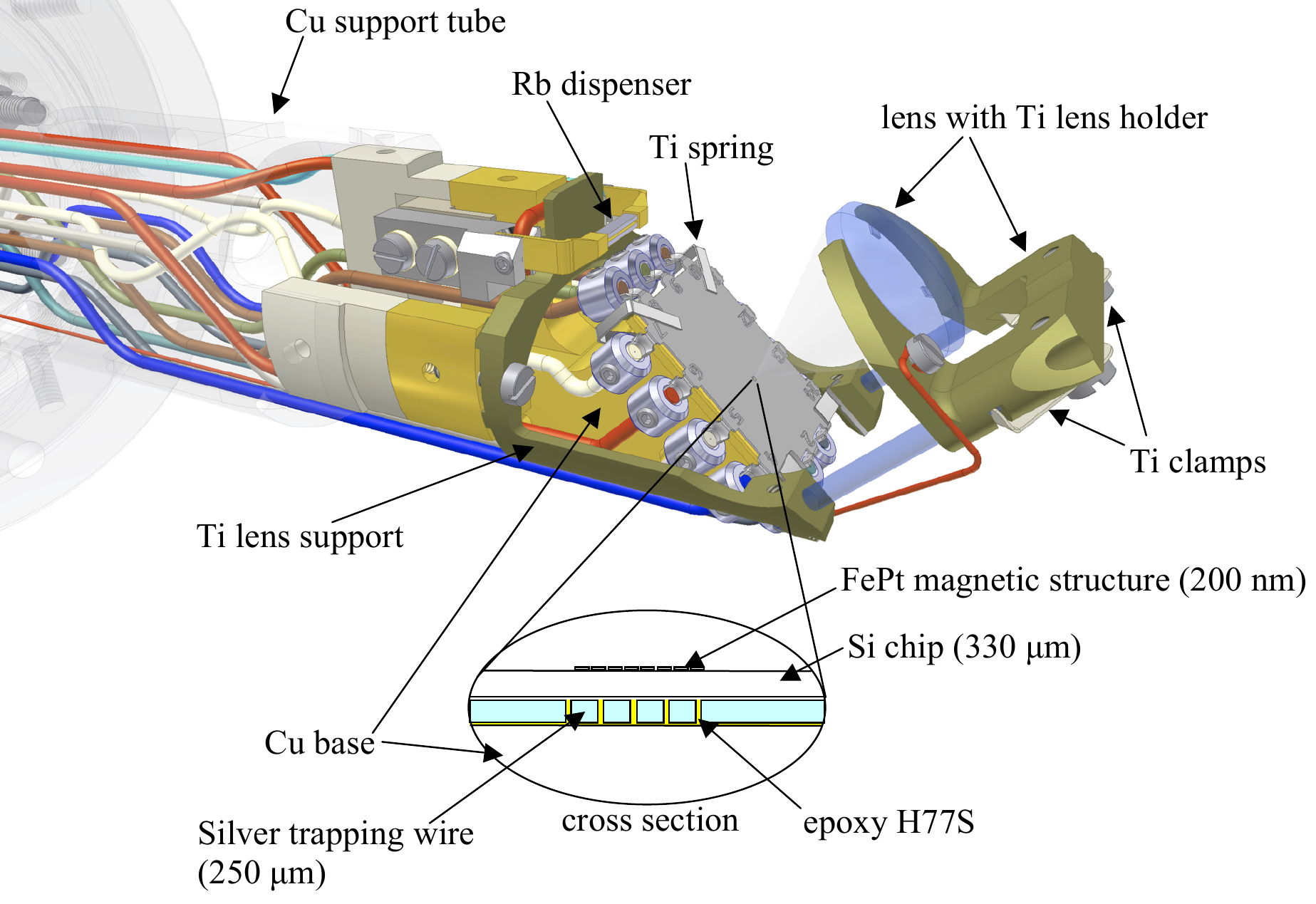}
    \caption{Schematic of design for the atom chip mount.}
    \label{schematic}
\end{figure}

The copper base carrying the atom chip is glued (Epotek E4110LV) to a copper support tube, providing a good thermal connection to the vacuum parts with high mechanical stability (Fig.~\ref{schematic}).  
The copper support tube is attached onto a reducing flange and attached to a CF63 cube.  The resulting structure containing the mounted chip is
finally  enclosed by a glass cuvette, which is  connected to the same reducing flange and serves as the experimental chamber.

\subsection{In-vacuum lens mount}

The chip mount and vacuum chamber are designed to allow for as much optical access as possible. Our starting point for the experiments is based on laser trapping and cooling $^{87}$Rb atoms in a standard mirror MOT configuration~\cite{ReiHanHan99}.  The mirror MOT uses four \unit[780]{nm} wavelength laser beams, with two counter-propagating beams reflected by the chip at a 45$^\circ$ angle. An extra \unit[780]{nm} beam is used for absorption imaging. For good optical imaging resolution, as well as addressing of single sites, an aspheric lens with numerical aperture NA=0.4 (Edmund Optics NT47-727, F=18.75 mm, D=15 mm) is mounted directly above the chip, inside the vacuum system. Based on these parameters  we expect a diffraction-limited optical resolution of \unit[1.2]{$\mu$m} (Rayleigh criterion), well below the trap lattice spacing of \unit[10]{$\mu$m}.  We also expect a detection sensitivity of approximately a single atom per site per shot \cite{OckTauSpr10,TauSpr13}.
The lens is held by a titanium lens holder, which is connected to the base mount by two quartz rods, as shown in Fig.~\ref{schematic}. The lens holder is fastened to the rods by two clamps, relying only on the mechanical flexibility of the clamps in order to minimize stress on the quartz rods and to prevent them from breaking. The clamps, lens support and lens holder are all made of titanium to minimize influences on the magnetic field. The lens surface facing the chip was coated with Indium Tin Oxide (ITO) and a dielectric anti-reflection coating.  The ITO coating is electrically connected to one of the pins in the electrical feedthrough and to the lens holder by a droplet of conducting Epotek E4110LV glue.
This can be used to apply an electric field using the lens surface as an electrode and the chip surface as a (ground) counter electrode, as explained in Section~\ref{layerCoating}. The ability to create an electric field gradient above the chip is an important tool for the future manipulation of Rydberg atoms.

\begin{figure}[h]
    \centering
    \includegraphics[width=1.5in] {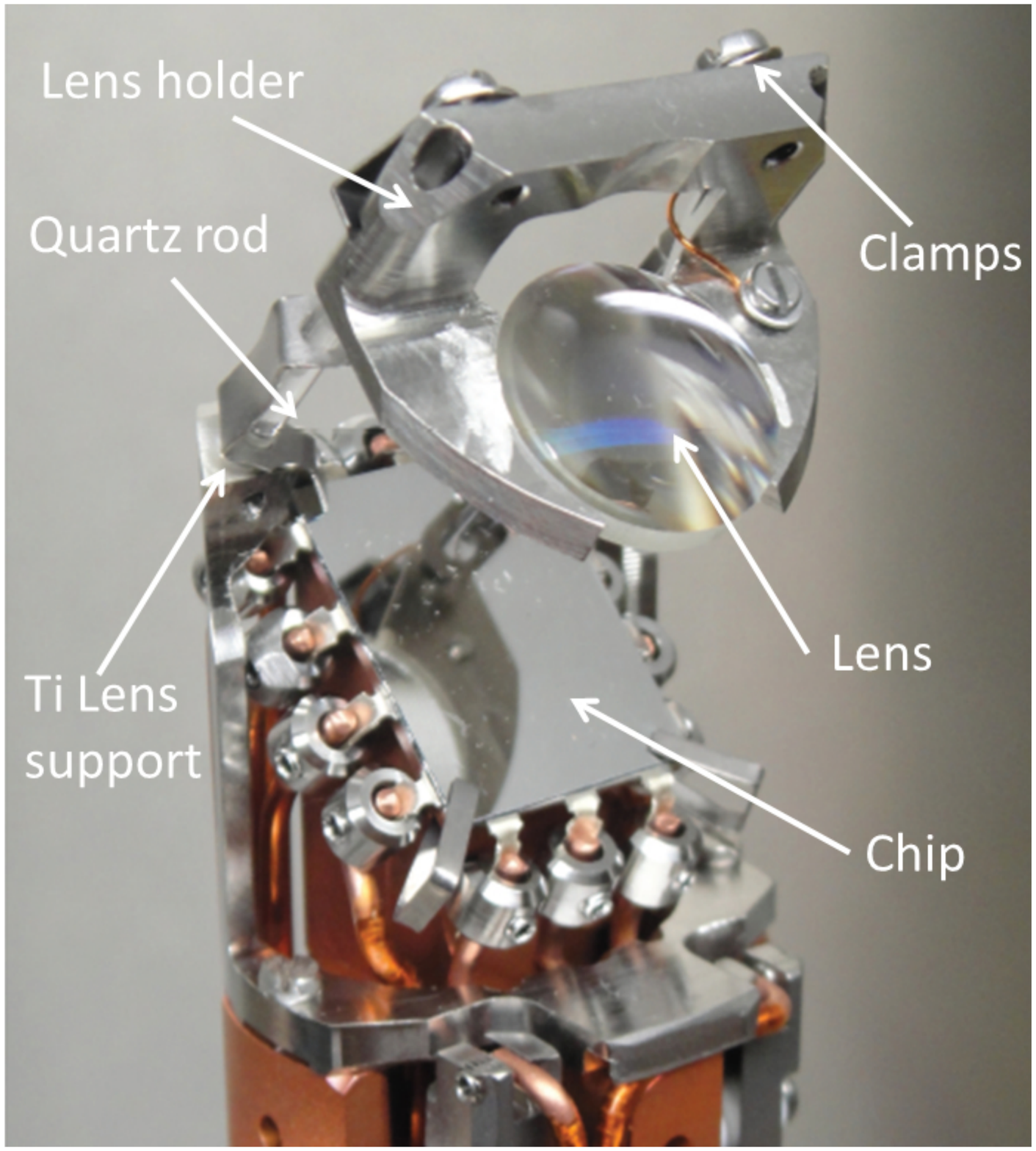}
    \caption{Mounted atom chip and in-vacuum lens attached to the Ti-lens holder.}
    \label{lensholder}
\end{figure}

Focusing the lens on the atoms is done by un-tightening the screws holding the clamps, and gently tapping on the lens holder. A laser beam passing through the lens is reflected by the chip and the outcoming light is compared to a reference beam traveling an equal distance (to correct for the natural divergence of the beam). The reference and chip beams are projected onto a CCD and the 1/e$^{2}$ size is compared. When the reference and chip beams are of equal size and divergence, the lens is focused on the chip. We started with the lens too close to the chip, giving a strongly divergent beam. By tapping from this side towards equally sized beams we avoid a focus of the beam in between the camera and the chip. The depth of focus for the chosen optics is \unit[5.2]{$\mu$m}. Since the atoms are trapped at \unit[$\sim$7]{$\mu$m} from the chip surface we cannot bring both the atoms and their reflections into the focus simultaneously. We chose to focus on the surface of the chip and use outside lenses for optional corrections. Beam divergence indicates that we have moved the focal point of the lens to within \unit[15]{$\mu$m} from the surface of the chip. With a tapping precision of about \unit[20]{$\mu$m}, this was accepted as optimal. After focusing, the clamps are tightened and the cuvette is carefully positioned over the entire assembly. To ensure that the lens stays in focus during the bakeout procedure, several thermal cycling tests have been performed. The imaging system is completed with extra lenses outside the vacuum system, to focus the image onto a CCD camera, and to adjust the magnification. Final focusing is done by moving the outside lenses and the position of the CCD camera.

\subsection{Vacuum setup}

The vacuum system for our experiment is based on a stainless steel (SS 304) cube containing six CF-63 ports. Opposite to the glass cuvette containing the chip, a 20 pin/\unit[10]{A} feedthrough is mounted for the electrical connections to the silver wires and the dispenser. The other four ports are used to connect the ion getter pump, the ion gauge, the turbo pump valve and an extra 4-pin feedthrough for the electric field connections and \emph{in vacuo} thermocouple. All materials used in the vicinity of the atom chip have been carefully chosen to be nonmagnetic, including: copper, silver, titanium, chromium, Kapton-insulated copper wires, Epotek glue, a rubidium dispenser and SS 316 steel screws.

A bakeout procedure is required in order to reach ultra high vacuum pressures. Bakeout temperatures are limited to \unit[150]{$^\circ$C} in order to prevent loss of magnetization of the FePt structure on the chip. After initial pumping with a turbo pump, a titanium-sublimation pump, and an ion getter pump, we gradually heat the whole system to \unit[150]{$^\circ$C}. After pumping for a week and subsequent cooling, a base pressure of \unit[$1\times 10^{-11}$]{mbar} has been achieved. As the atomic source we use enriched rubidium dispensers (Alvatec, s-type) to minimize the required gas load while achieving sufficient vapor pressure for a MOT.  The dispenser is located along one of the short edges of the chip and is used in pulsed-mode \cite{ForGrosHan98} to extend the lifetime of the atoms in the trap as we use a single-chamber.

\section{Loading atoms into the microtraps}

After the bakeout and alignment, we initially load $7\times 10^{7}$ atoms of $^{87}$Rb into the MOT using the dispenser in pulsed-mode. The initial cloud is then positioned using the U-wire, and cooled to \unit[50]{$\mu$K} using optical molasses.  After optical molasses, the cooled atomic cloud is optically pumped into the $|F =2, m_{F} = 2\rangle$ state.  The atoms are then loaded into the magnetic trap using the Z-wire of the chip, in combination with external magnetic coils.

Once in the magnetic trap, the atomic cloud is further compressed and cooled to \unit[10]{$\mu$K} by performing RF-evaporation. In order to load atoms into the microtrap array, the atomic cloud needs to be brought closer to the chip surface. This can be done by decreasing the current through the Z-wire $I_Z$, and/or increasing the external bias field component $B_y$ perpendicular to the central Z-wire section. Both methods compress the atomic cloud and push it towards the surface. Starting at a distance of approximately \unit[600]{$\mu$m} from the surface we compress the trap from a radial trap frequency of $\omega = 2\pi\times$~\unit[200]{Hz} to $2\pi\times$~\unit[800]{Hz} and move the trap to a distance of approximately \unit[200]{$\mu$m} from the chip.  Typical trap parameters at this point in the loading sequence are: $I_Z=\unit[17.5]{A}$, $B_x=\unit[-1.5]{G}$, and $B_y=\unit[50]{G}$.

\begin{figure}[h]
    \centering
    \includegraphics[width=3.5in]{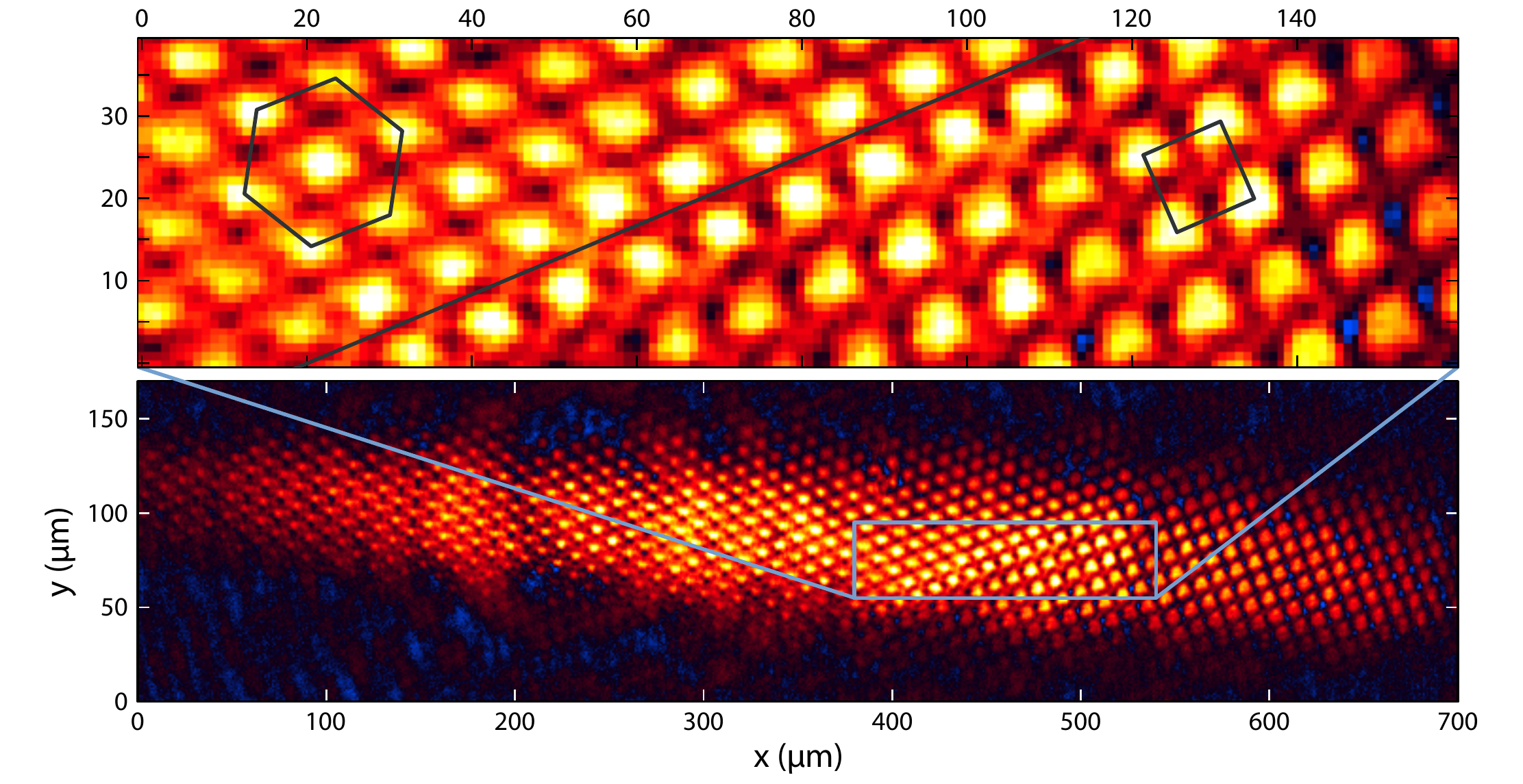}
    \caption{Absorption imaging of the loaded magnetic lattice. The rubidium atoms are loaded simultaneously into hexagonal (left side) and square (right side) lattices. A magnified view of the center region with the crossover between both geometries is shown in the upper part.}
    \label{microtraps}
\end{figure}

To lower the atomic cloud into the microtraps we turn off the current through the Z-wire and ramp the bias fields to $B_x=\unit[2]{G}$ and $B_y=\unit[5]{G}$. The microtraps are located at a distance to the chip of \unit[6.6]{$\mu$m} with a calculated radial trap frequency of $\omega = 2\pi\times$~\unit[25]{kHz} for the square lattice and $2\pi\times$~\unit[40]{kHz} for the hexagonal lattice. The component of the external field along the quantization axis of the IP field at the center of the atomic cloud is changed during the loading procedure, to prevent this component from crossing zero field. This turns out to be crucial to prevent spin-flips and thereby loss of a large fraction of the trapped atoms. An absorption image of a successfully loaded lattice of microtraps is shown in Fig.~\ref{microtraps}.

The atom chip has active trapping regions containing both square and hexagonal lattice geometries. These are clearly distinguishable in Fig.~\ref{microtraps} where the hexagonal lattice is on the left hand side and the square lattice on the right hand side. Initial analysis has been done to determine the atom number and temperature; these values are dependent on the bias field. Based on 41 absorption images and on the analysis of 175 (435) traps for the square (hexagonal) lattices respectively, we determine that we have on average $\sim$360 atoms per site in the square region and $\sim$440 atoms in the hexagonal region. {\em In situ} temperature measurements based on RF-induced atom loss spectroscopy show average temperatures of \unit[35$\pm$6]{$\mu$K} and \unit[32$\pm$5]{$\mu$K} for the square and the hexagonal regions, respectively. We find that the offset field at the trap bottoms is \unit[2.72(4)(16)]{G} for the square and \unit[1.97(3)(14)]{G} for the hexagonal region, where the parentheses present the fitting error and the trap-to-trap variation, respectively.

\section{Conclusion and Outlook}

We have described the design and construction of an advanced atom chip based on a lattice of permanent magnetic microtraps for neutral atoms.  Considerations which we have taken into account include: the electrical current load and thermal load of the preliminary magnetic trap based on current conducting wires, the possibility for rapid chip exchange, the mechanical stability of the chip, compactness and space limitations, optical accessibility and the requirement to image with a resolution in the micrometer range, and the needs of ultra-high vacuum. Following the construction and initial testing, the new atom chip has been placed in vacuum and finally tested by loading rubidium atoms into the magnetic microtraps after a sequence of compression and cooling stages.
Having atoms in a \unit[10]{$\mu$m} spacing lattice on an atom chip will allow us to access the Rydberg dipole-dipole interaction regime required for quantum information experiments.

The same approach using magnetic lattices on a chip could also be used  for scaling down the lattice period to the \unit[100]{nm} range. The distance between the traps and the chip surface will scale down linearly with the lattice period. Nevertheless, we estimated in previous work that the microtraps will easily overcome the Van der Waals attraction to the surface \cite{LeuTauSpr11}. While no longer compatible with Rydberg excitation, these lattices would  push Hubbard models that are now studied in optical lattices into novel parameter regimes \cite{Gullans2012Nanoplasmonic,Romero-Isart2013}. In particular, the tunneling rates and interaction energies will be greatly increased compared to optical lattices. We also estimated that heating due to Johnson noise originating in the conducting surface \cite{HenPotWil99,HenPot01,SchRekHin05,LinTepVul04} should not be prohibitive, owing to the small thickness of the magnetic films and their low conductivity. The strong heating effects observed in ion traps \cite{Deslauriers2006Scaling} close to surfaces do not play a role either, since no radio-frequency fields are involved.

To reach smaller dimensions, we will need new materials and new methods of fabrication which include  magnetic films with smaller grain size. In addition it will be necessary to switch from UV optical lithography to e-beam lithography. The ultimate limits on permanent magnet atom chips will be determined to a large extent by advances in the fields of magnetic materials and magnetic storage.

Several exciting directions for the development of atom chips based on permanent magnetic films lie ahead.  Technological advances such as multi-layering can be exploited to enable single site addressability with large parallelism.  An atom chip of multi-layers will allow the incorporation of a layer of electrodes, which can be fabricated from a low conductance material with reduced Johnson noise such as ITO.  This electrode layer will enable single site addressability by having tiny local electric fields act on the atom, taking it in and out of resonance with RF, microwave, or light radiation.  The ability to address a single atom within a large-scale two-dimensional atomic register enables the demonstration of algorithms which have so far been out of reach for both neutral atom and ion systems, and stands at the foundation of digital quantum simulation.

\section{Acknowledgements}

The authors would like to thank Anne de Visser for help in magnetizing the chip, Mattijs Bakker and the Technology Centre of the UvA Faculty of Science for their support.
RF gratefully acknowledges support by the Miller Institute for Basic Research in Science, University of California at Berkeley.
This work is part of the research program of
the Foundation for Fundamental Research on Matter (FOM),
which is part of the Netherlands Organization for Scientific
Research (NWO). The UvA authors acknowledge support from the EU
Marie Curie ITN COHERENCE Network.


\begin{thebibliography}{99}


\bibitem {Endres2011Observation}
	M.\ Endres, M.\ Cheneau, T.\ Fukuhara, C.\ Weitenberg, P.\ Schau{\ss}, C.\ Gross, L.\ Mazza, M.~C.\ Ba\~{n}uls, L.\ Pollet, I.\ Bloch, and S.\ Kuhr, Science {\bf 334}, 200 (2011).
	
\bibitem {Greif2013ShortRange}
	D.\
  {Greif},  {{T.\ Uehlinger}},
   {{G.\ Jotzu}},
   {{L.\ Tarruell}}, \ and\
   {{T.\ Esslinger}},
  {Science}\textbf { {340}},\  {1307}
  ({2013}).


\bibitem {Bloch2012Quantum}
	{{I.\
  Bloch}},  {{J.\ Dalibard}}, \
  and\  {{S.\ Nascimbene}},
   {Nat.\ Phys.}\ \textbf { {8}},\
  {267} ({2012}).

\bibitem {Simon2011Quantum}
	{J.\
  {Simon}},  {{W.~S.}\  {Bakr}},
   {{R.\ Ma}},
  {{M.~E.}\  {Tai}},  {
  {P.~M.}\  {Preiss}}, \ and\  {
  {M.\ Greiner}}, Nature {\bf 472}, {307} (2011).


\bibitem {GreManBlo02}
	{M.\
  {Greiner}},  {{O.\ Mandel}},
   {{T.\ Esslinger}},
   {{T.~W.}\  {H{\"a}nsch}}, \ and\
   {{I.\ Bloch}},\
  {   {   {Nature}\ }\textbf { {415}},\   {39} ({2002}).}

\bibitem {Seidelin2006}
   {S.\
  {Seidelin}},  {{J.\
  {Chiaverini}},  {{R.\ Reichle}},
   {{J.~J.}\  {Bollinger}},
   {{D.\ Leibfried}},
  {{J.\ Britton}},  {
  {J.~H.}\  {Wesenberg}},  {{R.~B.}\
   {Blakestad}},  {{R.~J.}\
   {Epstein}},  {{D.~B.}\
   {Hume}},  {{W.~M.}\
  {Itano}},  {{J.~D.}\  {Jost}},
   {{C.\ Langer}},
   {{R.\ Ozeri}},
  {{N.\ Shiga}}, \ and\
  {{D.~J.}\  {Wineland}},\ }{   {
  {Phys. Rev. Lett.}\ }\textbf {  {96}},\    {253003} ( {2006}).}

\bibitem {Blatt2008Entangled}
   {R.\
  {Blatt}}\ and\  {{D.\
  {Wineland}},\ } {
  {   {Nature}\ }\textbf {  {453}},\
  {1008} ({2008}).}

\bibitem {Islam2011Onset}%
   {R.\
  {Islam}},  {{E.~E.}\  {Edwards}},
   {{K.\ Kim}},
  {{S.\ Korenblit}},
  {{C.\ Noh}},  {
  {H.\ Carmichael}},  {{G.~D.}\
   {Lin}},  {{L.~M.}\
  {Duan}},  {{C.~C.}\
  {Joseph~Wang}},  {{J.~K.}\
  {Freericks}}, \ and\  {{C.\
  {Monroe}},\ }
  {   {Nat.\ Comm.}\ }\textbf {
  {2}},\  {377} ({2011}).


\bibitem {Lanyon2011Universal}%
   { {{B.~P.}\
  {Lanyon}},  {{C.\ Hempel}},
   {{D.\ Nigg}},
  {{M.\ M\"{u}ller}},
  {{R.\ Gerritsma}},
  {{F.\ Z\"{a}hringer}},
  {{P.\ Schindler}},
  {{J.~T.}\  {Barreiro}},
  {{M.\ Rambach}},  {
  {G.\ Kirchmair}},  {
  {M.\ Hennrich}},  {
  {P.\ Zoller}},  {
  {R.\ Blatt}}, \ and\  {{C.~F.}\
   {Roos}},\ } {
   {   {Science}\ }\textbf {
  {334}},\ {57} ({2011}).}


\bibitem{WelBauAba11}
J.\ Welzel, A.\ Bautista-Salvador, C.\ Abarbanel, V.\ Wineman-Fisher, C.\ Wunderlich, R.\ Folman and F.\ Schmidt-Kaler, Eur.\ Phys.\ J.\ D {\bf 65}, 285 (2011).

\bibitem{Piotrowicz2013Two-dimensional}
M.\ J.\ Piotrowicz and M.\ Lichtman and K.\ Maller and G.\ Li and S.\ Zhang and L.\ Isenhower and M.\ Saffman, Phys.\ Rev.\ A {\bf 88}, 013420 (2013).

\bibitem{Nogrette2014Singleatom}
F.\ Nogrette, H.\ Labuhn,  S.\ Ravets,  D.\ Barredo,  L.\ B\'{e}guin, A.\ Vernier, T.\ Lahaye, and A.\ Browaeys, 
arxiv:1402.5329 (2014).

\bibitem{JakCirLuk00}
D.\ Jaksch, J.~I.\ Cirac, P.\ Zoller, S.~L.\ Rolston, R.\ C\^{o}t\'{e}, and M.~D.\ Lukin, Phys.\ Rev.\ Lett.\
  {\bf 85}, 2208 (2000).

\bibitem{LukFleZol01}
M.~D.\ Lukin, M.\ Fleischhauer, R.\ C\^{o}t\'{e}, L.~M.\ Duan, D.\ Jaksch, J.~I.\ Cirac, and P.\ Zoller, Phys.\ Rev.\ Lett.\
  {\bf 87}, 037901 (2001).

\bibitem{WeiMulBuc10}
H.\ Weimer, M.\ M\"{u}ller, I.\ Lesanovsky, P.\ Zoller, and H.~P.\ B\"{u}chler, Nat.\ Phys.\
  {\bf 6}, 382 (2010).

\bibitem{WilGaeBro10}
T.\ Wilk,  A.\ Ga\"{e}tan, C.\ Evellin, J.\ Wolters, Y.\ Miroshnychenko, P.\ Grangier, and A.\ Browaeys, Phys.\ Rev.\ Lett.\
  {\bf 104}, 010502 (2010).

\bibitem{IseUrbSaf10}
L.\ Isenhower, E.\ Urban, X.~L.\ Zhang, A.~T.\ Gill, T.\ Henage, T.~A.\ Johnson, T.~G.\ Walker, and M.\ Saffman, Phys.\ Rev.\ Lett.\
  {\bf 104}, 010503 (2010).

\bibitem {Saffman2010Quantum}%
  M.\ Saffman, T.~G.\ Walker, and K.\ M{\o}lmer, Rev.\ Mod.\ Phys.\ {\bf 82}, 2313 (2010).

\bibitem{WhiGerSpr09}
S.\ Whitlock, R.\ Gerritsma, T.\ Fernholz, and R.\ J.\ C.\ Spreeuw, New J.\ Phys.\
  {\bf 11}, 023021 (2009).

\bibitem{GerWhiFer07}
R.\ Gerritsma, S.\ Whitlock, T.\ Fernholz, H.\ Schlatter, J.\ A.\ Luigjes, J.-U.\ Thiele, J.\ B.\ Goedkoop and R.\ J.\ C.\ Spreeuw, Phys.\ Rev.\ A
  {\bf 76}, 033408 (2007).


\bibitem{SinVolAku08}
M.\ Singh,  M.\ Volk, A.\ Akulshin, A.\ Sidorov, R.\ McLean, and P.\ Hannaford, J.\ Phys.\ B {\bf 41}, 065301 (2008).

\bibitem {LlorenteGarcia2010Experiments}%
 {I.\ Llorente~Garc\'{\i}a}, B.\ Darqui\'{e}, E.~A.\ Curtis, C.~D.~J.\ Sinclair,  and E.~A.\ Hinds, New J.\ Phys.\ {\bf 12}, 093017 (2010).

\bibitem {Boyd2007Atom}%
  M.\ Boyd, E.~W.\ Streed, P.\ Medley, G.~K.\ Campbell, J.\ Mun, W.\ Ketterle,  and D.~E.\ Pritchard, Phys.\ Rev.\ A {\bf 76}, 043624 (2007).

 \bibitem{LeuTauSpr11}
V.\ Y.\ F.\ Leung, A.\ Tauschinsky, N.\ J.\ van Druten, R.\ J.\ C.\ Spreeuw, Quantum Inf.\ Process.\
  {\bf 10}, 955--974 (2011).

\bibitem{Gullans2012Nanoplasmonic}
M.\ Gullans, T.\ G.\ Tiecke, D.\ E.\ Chang, J.\ Feist, J.\ D.\ Thompson, J.\ I.\ Cirac, P.\ Zoller, and M.\ D.\ Lukin, Phys.\  Rev.\  Lett.\ {\bf 109}, 235309 (2012).

\bibitem{Romero-Isart2013}
O.\ Romero-Isart, C.\ Navau, A.\ Sanchez, P.\ Zoller, and J.\ I.\ Cirac, Phys.\ Rev.\ Lett.\ {\bf 111}, 145304 (2013).

\bibitem{DitKoeHer08}
C.\ S.\ E.\ van Ditzhuijzen, A.\ F.\ Koenderink, J.\ V.\ Hernandez, F.\ Robicheaux, L.\ D.\ Noordam, and H.\ B.\ van Linden van den Heuvel, Phys.\ Rev.\ Lett.\
{\bf 100}, 24 (2008).

\bibitem{TauThiSpr10}
  A.\ Tauschinsky, R.\ M.\ T.\ Thijssen, S.\ Whitlock, H.\ B.\ van Linden van den Heuvell, and R.\ J.\ C.\ Spreeuw, Phys.\ Rev.\ A
  {\bf 81}, 063411 (2010).

\bibitem{Abel2011Electrometry}
R.\ P.\ Abel, C.\ Carr, U.\ Krohn, and C.\ S.\ Adams,
Phys.\ Rev.\  A {\bf 84}, 023408 (2011). 

\bibitem{Kubler2010Coherent}
H.\ K\"{u}bler, J.\ P.\ Shaffer, T.\ Baluktsian, R.\ L\"{o}w, and T.\ Pfau, Nat.\ Photonics {\bf 4}, 112 (2010)

\bibitem{Hattermann2012Detrimental}
H.\ Hattermann, M.\ Mack, F.\ Karlewski, F.\ Jessen, D.\ Cano, and J.\ Fort\'{a}gh, Phys.\  Rev.\  A {\bf 86}, 022511 (2012)


\bibitem{XinBarGer07}
 Y.\ T.\ Xing, I.\ Barb, R.\ Gerritsma, R.\ J.\ C.\ Spreeuw, H.\ Luigjes, Q.\ F.\ Xiao, C.\ R\'{e}tif, and J.\ B.\ Goedkoop, J.\ Magn.\ Magn.\ Mater.\
{\bf 313 }, 192-197 (2007).

\bibitem {Barmak2005}%
   {{K.\   {Barmak}},  {{J.\ Kim}},   {{D.~C.}\  {Berry}},   {{W.~N.}\  {Hanani}},
  {{K.\ Wierman}},  {  {E.~B.}\  {Svedberg}}, \ and\  {  {J.~K.}\  {Howard}},\ }    {   {J.\ Appl.\ Phys.}\ }\textbf { {97}}
  ({2005}).

\bibitem{SchSprWhi10}
R.\ Schmied, D.\ Leibfried, R.\ J.\ C.\ Spreeuw, and S.\ Whitlock, New J.\ Phys.\
  {\bf 12}, 103029 (2010).

\bibitem{RaaPrenCab87}
E.\ L.\ Raab, M.\ Prentiss, A.\ Cable, S.\ Chu and D.\ E.\ Pritchard, Phys.\ Rev.\ Lett.\  {\bf 59}, 2631 (1987).

\bibitem{ReiHanHan99}
J.\ Reichel, W.\ H\"{a}nsel, and T.\ W.\ H{\"a}nsch, Phys.\ Rev.\ Lett.\
 {\bf 83}, 3398 (1999).

\bibitem{BoFenMin09}
Y.\ Bo, C.\ Feng, K.\ Min, L.\ Xiao-Lin, T.\ Jiu-Yao and W.\ Yu-Zhu, Chin.\ Phys.\ B
{\bf 18}, 4259 (2009).

\bibitem{KleArl06}
C.\ Klempt, T.\ van Zoest, T.\ Henninger, O.\ Topic, E.\ Rasel, W.\ Ertmer, and J.\ Arlt, Phys.\ Rev.\ A
  {\bf 73 (1)}, 013410 (2006).

 \bibitem{GozBer93}
 A.\ Gozzini, F.\ Mango, J.\ H.\ Xu, G.\ Alzetta, F.\ Maccarrone, and R.\ A.\ Bernheim, Il Nuovo Cimento D
  {\bf 15 (5)}, 709--722 (1993).

\bibitem{FolFruScht02}
  R.\ Folman, P.\ Kruger, J.\ Schmiedmayer, J.\ Denschlag, and C.\ Henkel,  Adv.\ At.\ Mol.\ Opt.\ Phys.\
  {\bf 48}, 263-356 (2002).

   \bibitem{DuReiImi04}
S.\ Du, M.\ B.\ Squires, Y.\ Imai, L.\ Czaia, R.\ A.\ Saravanan, V.\
Bright, J.\ Reichel, T.\ W.\ H{\"a}nsch, and D.\ Z.\ Anderson, Phys.\ Rev.\ A
 {\bf 70}, 053606 (2004).

 \bibitem{OckTauSpr10}
C.\ F.\ Ockeloen, A.\ F.\ Tauschinsky, R.\ J.\ C.\ Spreeuw, S.\ Whitlock, Phys.\ Rev.\ A
  {\bf 82}, 061606(R) (2010).

\bibitem{TauSpr13}
A.\ Tauschinsky and R.\ J.\ C.\ Spreeuw, Opt.\ Ex.\
{\bf 21}, 10188 (2013).

 \bibitem{ForGrosHan98}
J.\ Fortagh and A.\ Grossmann, and T.\ W.\ H{\"a}nsch, J.\ Appl.\ Phys.\
 {\bf 84}, 12 (1998).

\bibitem{HenPotWil99}
C.\ Henkel, S.\ P\"{o}tting, and M.\ Wilkens, Appl.\  Phys.\ B
  {\bf 69}, 379 (1999).

\bibitem{HenPot01}
C.\ Henkel, and S.\ P\"{o}tting, Appl.\ Phys.\ B
  {\bf 72}, 73 (2001).

\bibitem{SchRekHin05}
S.\ Scheel, P.\ K.\ Rekdal, P.\ L.\ Knight, and E.\ A.\ Hinds, Phys.\ Rev.\  A
  {\bf 72}, 042901 (2005).

\bibitem{LinTepVul04}
Y.\ J.\ Lin,  I.\ Teper, C.\ Chin, and V.\ Vuletic, Phys.\  Rev.\  Lett.\ 
  {\bf 92}, 050404 (2004).

\bibitem{Deslauriers2006Scaling}
L.\ Deslauriers, S.\ Olmschenk, D.\ Stick, W.\ K.\  Hensinger, J.\ Sterk,  and C.\ Monroe, Phys.\  Rev.\  Lett.\ {\bf 97}, 103007 (2006).






\end{thebibliography}
\end{document}